\documentclass[%
prd
 ,secnumarabic%
,amssymb, amsmath,nobibnotes,showpacs]{revtex4}
\usepackage{amssymb,amsmath,amscd}
\usepackage{graphicx,calc,epsfig,pstricks,bbm}
\usepackage{tikz}

\begin{document}

\title{Dirac prescription from BRST symmetry in FRW space-time}
\author{Francesco Cianfrani}
\email{francesco.cianfrani@ift.uni.wroc.pl}
\affiliation{Instytut Fizyki Teoretycznej, Uniwersytet Wroc\l awski, pl. M. Borna 9, 50-204 Wroc\l aw, Poland, EU.}
\author{Giovanni Montani}
\email{giovanni.montani@frascati.enea.it}
\affiliation{ENEA - C.R, UTFUS-MAG, Via Enrico Fermi 45, 00044 Frascati, Roma, Italy, EU}
\affiliation{Dipartimento di Fisica, Universit\`a di Roma ``Sapienza'', Piazzale Aldo Moro 5, 00185 Roma, Italy, EU}

\begin{abstract}
A procedure to define the BRST charge from the Noether one in extended phase space is given. It is outlined how this prescription can be applied to a Friedmann-Robertson-Walker space-time with a differential gauge condition and it allows us to reproduce the results of \cite{Tatyana3}. Then we discuss the cohomological classes associated with functions in extended phase space having ghost number one and we recover the frozen formalism for classical observables. Finally, we consider the quantization of BRST-closed states and we define a scalar product which implements the superHamiltonian constraint.  
\end{abstract}


\maketitle

\section{Introduction}

The realization of a quantum theory for the gravitational field represents one of the most challenging issue in theoretical physics. In view of the up-to-now lack of experimental data which can guide us in the realization of a final theory, the current attempts (as for instance Loop Quantum Gravity \cite{lqg} and the Asymptotic Safety Scenario in Quantum Gravity \cite{asg}) are based on extending to geometrodynamics some ideas (as quantization of the holonomy-flux algebra and Wilsonian renormalization approach, respectively) which have found a fruitfull application in the description of fundamental interactions. In this work, we will take BRST symmetry as our guide to analyze the fate of time parametrization invariance in a quantum theory for the Friedmann-Robertson-Walker model.  

BRST simmetry \cite{brst} plays a prominent role in the standard paradigm of fundamental interactions. In fact, the development of a meaningful path-integral formulation for a Yang-Mills theory requires to fix a gauge condition, which implies that the original gauge symmetry is broken. Nevertheless, the extension of phase-space via the introduction of additional variables allows us to recover a formulation with a residual global invariance, {\it i.e.} BRST symmetry, associated with nihilpotent transformations. It is such a symmetry which implies that transition amplitudes do not depend on the adopted gauge-fixing condition, such that all the nice properties ({\it in primis} renormalization) found through a perturbative expansion hold on a non-perturbative level too \cite{Weinberg}.

Moreover, it is possible to trace back the origin of BRST symmetry to the peculiar features of the phase-space for constrained systems \cite{ht}. In fact, in the presence of first-class constraints observables are defined by a two-step procedure: i) they must be restricted to the constraint hypersurfaces, ii) they must be constant along the gauge orbits. These two requirement can be satisfied by enlarging phase-space via the introduction of some Grassmanian variables and by defining observables as functions belonging to the first cohomological class of a differential operators $s$ implementing i) and ii). It is possible to associate a canonical action to $s$, such that a proper BRST charge $\Omega$ can be defined. This is the case on a Hamiltonian level, the Batalin-Fradkin-Vilkowsky (BFV) \cite{BFV} theory, as well as on a Lagrangian level, the Batalin-Vilkovinsky (BV) \cite{BV} framework. This two formulations have been proved to be perturbatively equivalent \cite{bt}. However, in general $s$ contains some additional higher order (in ghost number) terms, such that one must consider an iterative expansion in the ghost number and look for a solution order by order (see for instance \cite{dh}). Only if the gauge symmetry is simple (as in the case of a Lie algebra) an explicit expression is known for the charge. 

For gravity in a 3+1 representation, the gauge group is the product of 3-diffeomorphisms and time reparametrizations. The gauge algebra is open and the definition of a proper BRST charge has still technical issues, such that it has been accomplished only on a perturbative level \cite{fai} or in 2+1 dimensions \cite{gp}. The applications of the BFV formalism in minisuperspace has been realized in the seminal work of Halliwell \cite{Hall}, where the Wheeler-DeWitt equation has been recovered from a path integral formulation.

In \cite{Tatyana1} the BV approach has been adopted to develop a proper BRST invariant Lagrangian for a Bianchi IX model in the presence of a differential gauge condition. Thanks to the presence of time derivatives of the lapse function in the total Lagrangian, the Hamiltonian in extended phase space is free from constraints and it can be defined simply by a Legendre transformation. By the direct inspection of the equations of motion, it was inferred the expression of the BRST charge in a Bianchi IX model. However, the definition of the BRST charge and transformations for more general metrics is hampered by the fact that the equations of motion are too complex for a direct inspection. Then, it has been suggested that BRST symmetry does not hold on a quantum level in absence of asymptotic states (as for a closed FRW model). This lead to the materialization of a reference frame, whose cosmological implications have been discussed in \cite{tcosm}.

In this work, starting from the Hamiltonian formulation developed in \cite{Tatyana1}, we outline how to find out the proper BRST charge. In a theory as GR, one deals with Lagrangian containing second order derivatives, which can be avoided by perfoming some partial integrations and discarding boundary contributions. Of course, this procedure does not affect the classical equations of motion, which are evaluated by considering vanishing variations at the boundary, thus also the classical symmetries are untouched. However, the resulting action can be not explicitly invariant under the original symmetries and the variation can provide some boundary contributions. These boundary contributions will enter the definition of the conserved charge.  
We will show that this is the case for a FRW model, 
in which the first order action is not BRST-invariant, but a boundary contribution arises. By properly accounting for these additional terms, the conserved charge can defined and its expression coincide with the one given in 
\cite{Tatyana1}. 

Then, we will characterize the BRST cohomological classes for functions having ghost number one. We will show how it is possible to choose proper functions $\Phi_{rep}$ along the orbits generated by the BRST charge such that the closure condition fixies the dependence from the lapse function, while exact forms are obtain via the Poisson action of the superHamiltonian. This achievements ensure that a 1-1 correspondence between classical observables and BRST cohomological classes exists and that the Poisson action of the physical Hamiltonian in extended phase space vanishes. 
Finally, we will quantize the system by considering $\Phi_{rep}$ as wave functions. We will find out a proper definition of the scalar product, which implements the superHamiltonian constraint mimicking the procedure defined in \cite{ht}. This way, after integrating out the ghosts and gauge variables, we will infer an expression for the scalar product in the kinematical Hilbert space which reproduces the result of refined algebraic quantization \cite{raq1,raq2}, thus also the Dirac prescription for the quantization of constrained systems \cite{dir}.     

In particular, the manuscript is organized as follows. In sec.\ref{sec1} we review the prescriptions given by Dirac, by refined algebraic quantization and by the BRST formulation for the canonical quantization of constrained systems. In sec.\ref{s1} the Hamiltonian formulation of the FRW model is presented in extended phase space. Sec. \ref{s2} is devoted to establish the relationship between the Noether charge and the one in the presence of boundary contributions, so giving a new derivation for the BRST charge in the FRW case. The cohomological classes are discussed in sec.\ref{s4}, while in sec.\ref{s4bis} the quantization of the associated system is defined and the scalar product implementing the superHamiltonian constraint is defined. Brief concluding remarks follow in sec.\ref{s5}. 

\section{Dirac prescription from BRST cohomological class}\label{sec1}

The observables in a theory with some first-class constraints $G_a(q,p)=0$ are defined as those phase-space functions $O(q,p)$ which are invariant under the Poisson action of the constraints, {\it i.e.} the relations $\{G_a,O\}=0$ hold.  The quantization of such a theory is based on the Dirac prescription \cite{dir}, {\it i.e.} the physical states $\psi_{phys}$ are those ones for which 
\begin{equation}
\hat{G}_a\psi_{phys}(q)=0,\label{dirp}
\end{equation}  

$\hat{G}_a$ being the operators associated with the phase-space functions $G_a$. 

An equivalent formulation on a quantum level is obtained in the so-called refined algebraic quantization \cite{raq1,raq2}, 
in which one works with generic states $\psi=\psi(q)$ and implements the condition $G_a=0$ in the scalar product as follows
\begin{equation}
<\psi_1,\psi_2>=\int dq\mu(q) \delta(G_a) \psi_1^*\psi_2.\label{raq}
\end{equation}

$\mu$ being a proper measure. 
In order to get a finite result for the scalar product (\ref{raq}), some gauge-fixing conditions $\chi_a=0$ have to be implemented via the insertion of $\delta$-functions and of some factors ensuring the invariance under the gauge choice in the measure $\mu$ \cite{ht,bi}. This can be an intriguing point. In what follows, we will concentrate on how to implement the restriction to the hypersurface where the constraints $G_a=0$ hold and we assume the $\delta$-functions of the gauge-fixing condition and the proper factors to be contained into the measure $\mu$.

Hennaux and Teitleboim \cite{ht} pointed out how the characteration of observales and of physical quantum states can be inferred from the BFV formulation in extended phase space. In particular, observables are elements of a proper BRST cohomological class, while it can be defined a scalar product which reproduces Eq.(\ref{raq}). 

In extended phase-space one introduces the ghosts $\theta^a$ and the antighosts $\bar\theta_a$ associated with the constraints $G_a=0$. Let us consider also the so-called nonminimal sector, in which one treats also the Lagrangian multipliers $\lambda^a$, implementing first-constraints in the Hamiltonian, on equal footing as others variables and their associated conjugate momenta $b_a$ are introduced. Hence, the additional conditions $b_a=0$ are present, together with the associated couple of ghosts-antighosts variables $\rho^a,\bar{C}_a$.

According with the BFV formulation \cite{BFV}, the total BRST charge reads
\begin{equation}
\Omega=-\theta^aG_a+\rho^ab_a+\bar\eta_aC^a_{bc}\theta^b\theta^c+\ldots,\label{bfvc}  
\end{equation}
where $\{G_a,G_b\}=C^c_{ab}G_c$, while $\ldots$ denotes some terms of higher order in the anti-ghost number of the minimal sector. 

Let us now consider the following functions of configuration variables having maximum ghost number in the minimal sector
\begin{equation}
\Phi=\phi(q,\lambda)\Pi_a\theta^a,\label{stat}
\end{equation}  

where the product extend over all the ghosts $\theta^a$, while $\phi(q)$ is a function of configuration variables only (it does not depend on $\lambda^a$). The crucial property of the functions (\ref{stat}) is the following 
\begin{equation}
\theta^a\Phi=0,\label{g0}
\end{equation}

which is due to the fact that $\Psi$ already contains all the available ghosts and $(\theta^a)^2=0$. The Poisson brackets with the charge gives 
\begin{eqnarray}
\{\Omega,\Phi\}=\{-\theta^aG_a,\Phi\}+\{\rho^ab_a,\Phi\}+\{\bar\theta_aC^a_{bc}\theta^b\theta^c+\ldots,\Phi\}=\nonumber\\
=-\theta^a\{G_a,\phi\}\Pi_b\theta^b+\rho^a\{b_a,\phi\}\Pi_b\theta^b+\theta^b\theta^c\{\bar\theta_aC^a_{bc},\Phi\}=-\rho^a\frac{\partial\phi}{\partial\lambda^a}\Pi_b\theta^b, 
\end{eqnarray}
 
where in the second line we used the relation (\ref{g0}). Henceforth, the functions (\ref{stat}) are BRST closed if $\phi$ does not depend on the Lagrangian multipliers $\lambda^a$, {\it i.e.}
\begin{equation} 
\Phi=\phi(q)\Pi_a\theta^a.\label{brstcl}
\end{equation}

Two different BRST-closed functions $\Phi_1=\phi_1\Pi_a\theta^a$ and $\Phi_2=\phi_2\Pi_a\theta^a$ belong to the same cohomological class ($\Phi_1=\Phi_2+\{\Omega,\Psi\}$) if there exists a third function of configuration variables only $\psi=\psi(q)$ such that
\begin{equation}
\phi_1=\phi_2+\{G_a,\psi\},
\end{equation} 

for some $a$. This means that each cohomological class $\{\phi\}$ is made of the gauge orbits of phase-space functions, {\it i.e.} 
\begin{equation}
[\phi]=\{\phi_1|\phi_1=\phi+\{G_a,\psi\}\}.\label{eqcl}
\end{equation}

It can be demonstrated that such cohomological classes are isomorphic to the cohomological classes of the states with minimum ghost number in the minimal sector, {\it i.e.}
\begin{equation}
\varphi=\varphi(q).\label{ming}
\end{equation}

The functions (\ref{ming}) are closed if 
\begin{equation}
\{\Omega,\varphi\}=\theta^a\{G_a,\varphi\}=0\rightarrow \{G_a,\varphi\}=0,\label{co}
\end{equation}

and they coincide with cohomological classes since no exact state of the kind (\ref{ming}) exists. Therefore, the elements of the BRST cohomological class $H^1(\Omega)$ are functions of the gauge orbits only, thus they are in 1-1 correspondence with the observables of the theory.


On a quantum level, the identification of proper quantum states can be done according with the procedure implemented in non-Abelian gauge theories \cite{Wein}. In fact, in such models the imposition of the invariance under the choice of the gauge fixing implies BRST invariance for amplitudes. These amplitudes are evaluated between asymptotic states, which are taken from free field theory. Hence, having proper in- and out- Hilbert spaces, the requirement of BRST invariance for amplitudes becomes a restriction of the space of admissible states, {\it i.e.} the space of BRST closed states. 

For gravitational systems like the FRW model, we do not have at our disposal a so-clear picture for quantization and in some cases ($k=1$) we cannot define asymptotic states at all. This feature leads to conjecture that BRST symmetry cannot be implemented on a quantum level \cite{Tatyana1}. Here we take the opposite point of view and starting from BRST closed states (\ref{brstcl}), we look for a proper definition of the scalar product implementing the restriction to the cohomology classes (\ref{eqcl}). This can be realized as follows \cite{ht} (square brackets denote the commutator)
\begin{equation} 
\langle\Phi_1,\Phi_2\rangle=\int dq\mu(q)\Pi_a d\lambda^ad\theta_ad\rho^a \Phi^*_1 e^{i[\hat{K},\hat{\Omega}]}\Phi_2,\label{sp}
\end{equation}
where $\hat{K}$ reads\footnote{In the following we will not put hats $\hat{}$ 
on multiplicative operators.}
\begin{equation}
\hat{K}=-i\lambda^a\hat{P}_a,\label{k}
\end{equation}

$\hat{P}_a$ being the operators associated with the conjugate momenta to $\theta^a$. In fact, one finds
\begin{equation}
[\hat{K},\hat{\Omega}]=\lambda^a\hat{G}_a-\hat{P}_a\rho^a+\ldots,
\end{equation}

such that the scalar product (\ref{sp}) becomes 
\begin{eqnarray}
\langle\Phi_1,\Phi_2\rangle=\int dq\mu(q)\Pi_a d\lambda^ad\theta_ad\rho^a \Phi^*_1 e^{i\lambda^a\hat{G}_a}\Pi_a(1-i\hat{P}_a\rho^a+\ldots)\Phi_2.\label{sp2}
\end{eqnarray}

The term $-i\Pi_a\hat{P}_a\rho^a$ transforms the ghosts $\Pi_a\theta^a$ inside $\Psi_2$ into $\Pi_a\rho^a$, so the Berezin integration over $d\theta_a d\rho^a$ gives non-vanishing finite results, while the integration over $\lambda^a$ provides the restriction to the subspace for which $G_a=0$, {\it i.e.}
\begin{equation}
\langle\Phi_1,\Phi_2\rangle=\int dq\mu(q)\phi^*_1(q)\delta(G_a)\phi_2(q).\label{spf}
\end{equation}

Therefore, the definition of the scalar product (\ref{sp}) provides the restriction to the constraint hypersurfaces where the constraints $G_a=0$ holds and it reproduces the results of the refined algebraic quantization (\ref{raq}) and of the Dirac prescription.  

This procedure is rather formal, because we do not specify the space where $\phi$'s live, the integration measure 
(which contains the $\delta$-functions over some gauge-fixing conditions) 
and the complex structure. We are going to apply it to the FRW case.

\section{BRST charge in FRW model}\label{s1}

The metric tensor for FRW models is given in spherical coordinates $\{t,r,\theta,\phi\}$ by 
\begin{equation}
ds^2=N^2dt^2-a^2(\frac{1}{1-kr^2}dr^2+r^2d\theta^2+r^2\sin^2\theta d\phi^2),
\end{equation}
$N$ and $a$ being the lapse function and the scale factor, respectively, which depend on the time variable $t$ only, while $k=1,0,-1$ for a closed, flat and open Universe, respectively. 

The Lagrangian density takes the following expression
\begin{equation}
L=-\frac{1}{2}\frac{a\dot{a}^{2}}{N}+\frac{k}{2}Na,\label{lgeo}\end{equation}
where $\dot{a}$ denotes the derivative of $a$ with respect to the time coordinate $t$. One sees that the conjugate momentum to the lapse function $N$ is constrained to vanish 
\begin{equation}
\pi=0. \label{1con}
\end{equation} 

By a Legendre transformation one finds the following expression for the Hamiltonian  
\begin{eqnarray}
{H}_{g}=\int N\mathcal{H}\label{h}d^3x,
\end{eqnarray}
in which the superHamiltonian $\mathcal{H}$ is given by
\begin{equation}
\mathcal{H}=-\frac{1}{2a}\pi_{a}^2-\frac{k}{2}a,\label{sham}
\end{equation}
$\pi_a$ being the conjugate momenta to $a$.

The conservation of the condition (\ref{1con}) implies the secondary constraint 
\begin{equation}
\mathcal{H}=0,\label{2con}
\end{equation}
such that the total Hamiltonian is constrained to vanish and no physical evolution occurs (frozen formalism). The observable of the theory are those phase space functions which commute with the constraints (\ref{1con}) and (\ref{2con}), {\it i.e.}
\begin{equation}
\{O,\pi\}=\{O,\mathcal{H}\}=0.\label{obsfrw}
\end{equation}

The presence of the constraints (\ref{1con}) and (\ref{2con}) is due to the fundamental gauge symmetry of the FRW dynamical system, which is the invariance under time parametrizations, {\it i.e.}
\begin{equation}
t'=t+\eta(t),\label{trp}
\end{equation}
where $\eta=\eta(t)$ is an infinitesimal parameter.

In order to give a well-defined formulation in extended phase space the following gauge condition has been considered in \cite{Tatyana3}
\begin{equation}
\dot{N}=\frac{dF}{da}\dot{a},\label{gcon} 
\end{equation}
such that, once the ghost $\theta$ and the antighost $\bar{\theta}$ are introduced, the Lagrangian density in extended phase space containing only first-derivatives reads
\begin{eqnarray}
L_{ext}=-\frac{1}{2}\frac{a\dot{a}^{2}}{N}+\frac{k}{2}Na+\lambda\left(\dot{N}-\frac{dF}{da}\dot{a}\right)+\dot{\overline{\theta}}\left(\dot{N}-\frac{dF}{da}\dot{a}\right)\theta+\dot{\overline{\theta}}N\dot{\theta}=\nonumber\\=-\frac{1}{2}\frac{a\dot{a}^{2}}{N}+\frac{1}{2}Na+\pi\left(\dot{N}-\frac{dF}{da}\dot{a}\right)+\dot{\overline{\theta}}N\dot{\theta},\label{lext}
\end{eqnarray}
$\lambda$ being a Lagrangian multiplier which imposes the gauge conditions (\ref{gcon}), while $\pi=\lambda+\dot{\bar{\theta}}\theta$. 

The analysis of equations of motion performed in \cite{Tatyana1} gives the following expression for the BRST charge 
\begin{equation}
\Omega=-H\theta-\pi\rho,\label{brstfrw}
\end{equation}
$\rho$ being the conjugate momentum to $\bar\theta$, while $H$ denotes the total Hamiltonian in extended phase space, which can be written as
\begin{equation}
H=N\widetilde{H}-\frac{\bar{\rho}\rho}{N},\label{ham}
\end{equation}
in which $\bar\rho$ denotes the conjugate momentum to $\theta$ and $\widetilde{H}$ can be obtained from $\mathcal{H}$ (\ref{sham}) by replacing $\pi_a$ with $\pi\frac{\partial F}{\partial a}+\pi_{a}$, {\it i.e.}
\begin{equation}
\widetilde{H}=-\frac{1}{2a}\left[\pi\frac{dF}{da}+\pi_{a}\right]^{2}-\frac{k}{2}a.\label{tildeh}
\end{equation}
  
It is worth noting the difference between the charge (\ref{brstfrw}) and the expression (\ref{bfvc}) obtained in the BFV model, in particular as soon as the Lagrangian multiplier $N$ and $\lambda$'s are concerned. This fact reflects the non-trivial mixing between gauge and physical degrees of freedom which takes place in the case of gravitational systems.  
 
\section{BRST charge}\label{s2}

Let us consider a dynamical system, described by some fields $\phi$ and a Lagrangian density $L(\phi,\partial\phi)$, and a group of  infinitesimal transformations  
\begin{equation}
x^\mu\rightarrow x'^{\mu}=x^{\mu}+\eta^\mu,\qquad \phi(x)\rightarrow \phi'(x)=\phi(x)+\delta\phi(x).
\end{equation}

The variation of the action under the transformations above, when evaluated on classical trajectories, reads
\begin{eqnarray}
\delta S=\int d^{4}x'L(\phi',\partial'\phi')-\int d^{4}xL(\phi,\partial\phi)=\nonumber\\=\int d^{4}x(1+\partial_{\mu}\eta^{\mu})L(\phi',\partial'\phi')-\int d^{4}xL(\phi,\partial\phi)=\nonumber\\
=\int d^{4}x\left[\frac{\delta L}{\delta\phi}\delta\phi+\frac{\delta L}{\delta\partial_\mu\phi}\delta\partial_\mu\phi+\partial_{\mu}\eta^{\mu}L+\eta^{\mu}\partial_{\mu}L\right]=\nonumber\\
=\int d^{4}x\partial_{\mu}\left(\frac{\delta L}{\delta\partial_{\mu}\phi}\delta\phi+\eta^{\mu}L\right),\label{var}
\end{eqnarray}

where in the last line a partial integration occurs and the Euler-Lagrange equations are used. The requirement of gauge invariance is usually implemented by imposing $\delta S=0$, from which one infers the conservation of the Noether charge $Q=\int\bigg[\frac{\delta L}{\delta\partial_{0}\phi}\delta\phi-\theta^{0}L\bigg]d^3x$ by discarding spatial boundary contributions.

However, it is possible to weaken such a condition and to require the variation of the action to be a boundary term, {\it i.e.} 
\begin{equation}
\delta S=-\int \partial_\mu D^\mu d^4x.\label{Dmu}
\end{equation}

The transformations for which the condition (\ref{Dmu}) holds are actual symmetries on a classical level. This is due to the fact that the equations of motion are evaluated by performing variations of the field which vanish at the boundary. Hence, a conserved charge $\Omega$ can still be defined and it differs from the expression $Q$ (which is not a Noether charge since the action is not invariant)
by a term $D^0$, {\it i.e.} 
\begin{equation}
\Omega=Q+\int D^0 d^3x. \label{omega}
\end{equation}
In fact, it can be verified from the relations (\ref{var}) and (\ref{Dmu}) that
\begin{equation}
\partial_t\Omega=\partial_tQ+\partial_t\int D^0 d^3x=\delta S-\delta S=0,
\end{equation}
where we discard the spatial boundary contributions. 

In the following, we will outline how this is the case for the BRST symmetry in FRW models and we will infer the expressions for the corresponding charge.

\subsection{FRW model} Let us now consider the FRW model. The BRST transformations associated with time translations (\ref{trp}) can be obtained by replacing the infinitesimal parameter $\eta$ with $b\theta$, $b$ being a constant Grassmanian parameter. The behavior of $N$ and $a$ is the following one
\begin{equation}
\delta N=-\dot{N}b\theta-Nb\dot{\theta},\qquad
\delta a=-\dot{a}b\theta.\label{1}
\end{equation}

The transformation of the ghost can be deduced from the fact that $\theta$ behaves as an infinitesimal displacement of the time coordinate, {\it i.e.} as the time-like component of a vector field. Hence, the variation of $\theta$ gives
\begin{equation}
\delta\theta=+b\dot{\theta}\theta,\label{2}
\end{equation}

while for $\lambda$ and $\bar{\theta}$ we fix in analogy with the BRST transformations in the Yang-Mills case, the following relations
\begin{equation}
\delta\overline{\theta}=-b\lambda,\qquad \delta\lambda=0.\label{3}
\end{equation}

It can be explicitly verified that the transformation defined by Eqs (\ref{1}), (\ref{2}) and (\ref{3}) is nihilpotent.

From the evaluation of the on-shell variation of the Lagrangian density (\ref{lext}), one finds that the variation of the gravitational part vanishes when the equations of motion for $N$ and $a$ hold, while the total variation reads
\begin{eqnarray}
\delta L=\delta(L_{gf}+L_{gh})=\delta\lambda(\dot{N}-\dot{F})+\lambda\partial_{t}\delta(N-F)-\nonumber\\-\delta\dot{\overline{\theta}}\delta(N-F)-\dot{\overline{\theta}}\delta\delta(N-F)=b\partial_{t}\left[\lambda\delta(N-F)\right]=b\partial_{t}\left[\lambda\rho\right].
\end{eqnarray}

The total variation of the action is given by summing to the variation above the term due to the transformation from $t'$ to $t$. This gives a contribution $\int \partial_0(\theta L)dt$, which evaluated on classical trajectories reduces to 
\begin{equation}
\int \partial_t(\theta L)dt=b\int \partial_t (\dot{\overline\theta}\theta\rho) dt, 
\end{equation}

Therefore the full variation of the action reads
\begin{equation}
\delta S=-b\int \partial_t (\pi\rho) dt. 
\end{equation}

Hence, the transformations (\ref{1}), (\ref{2}) and (\ref{3}) act as a symmetry for classical trajectories in extended phase space, because the variation of the action is a just boundary term which does not contribute to the equations of motions. Moreover, these transformations are nihilpotent and they constitutes the proper BRST transformations of the FRW model.  

At this point, the conserved charge can be evaluated from the Noether charge $Q=-\theta H-\pi N\dot{\theta}-\pi\rho$, by summing $\pi\rho$ so finding the following expression,
\begin{equation}
\Omega=-\theta H-\pi N\dot{\theta}.
\end{equation}

This expression coincides on-shell with the one obtained in \cite{Tatyana1} and, in fact, it generates the transformations (\ref{1}), (\ref{2}) and (\ref{3}). 
This result outlines how the direct evaluation of the on-shell variation of the action allows us to infer the right expression for the BRST charge $\Omega$ starting from the Noether charge $Q$.

\section{Cohomological classes}\label{s4}

Let us now discuss the cohomological classes of the BRST charge (\ref{brstfrw}) having ghost number one, $H^1(\Omega)$. 

In the FRW case, these functions in extended phase space can be written as
\begin{equation} 
\Phi(a,N,\theta,\rho)=\phi_\theta(N,a)\theta+\phi_\rho(a,N)\rho, \label{ffrw}
\end{equation}

$\phi_\theta$ and $\phi_\rho$ being arbitrary functions of $a$ and $N$.

The requirement of BRST invariance implies that
\begin{equation}
\{\Omega,\Phi\}=0\rightarrow \frac{\partial\phi_\theta}{\partial N}-\frac{\phi_\theta}{N}+N\{\widetilde{H},\phi_\rho\}=0.\label{inv}
\end{equation}

The exact functions $\Theta$ with total ghost number one can be obtained from a generic function $\varphi(a,N)$ as follows
\begin{equation}
\Theta=\{\Omega, \varphi\}=-N\{\widetilde{H},\varphi\}\theta-\frac{\partial}{\partial N}\varphi\rho,\label{exct}
\end{equation}

and two functionals $\Phi_1$ and $\Phi_2$ belong to the same orbit if there exists $\varphi$ such that $\Phi_1=\Phi_2+\{\Omega,\varphi\}$. 

Let us partially fix an element $\Phi_{rep}$ within each BRST orbit by the condition 
\begin{equation}
\phi_\rho=0, \label{repr}
\end{equation}

which can be realized starting from a generic function (\ref{ffrw}) by summing $\Theta=\{\Omega,\varphi\}$ with 
\begin{equation}
\frac{\partial \varphi}{\partial N}=-\phi_\rho \Rightarrow\varphi=-\int^N \phi_\rho(N',a)dN'. 
\end{equation} 
 
As soon as Eq.(\ref{repr}) holds, the relation (\ref{inv}) implies that BRST invariant functions read
\begin{equation}
\{\Omega,\Phi_{rep}\}=0\Rightarrow \Phi_{rep}=N\phi(a)\theta, \label{rinv}
\end{equation} 

$\phi(a)$ being an arbitrary function of the scale factor. The exact functions of the kind (\ref{rinv}) can be obtained by the Poisson action of the charge $\Omega$ on the ghost zero functions
\begin{equation}
\widetilde\varphi=\varphi(a)-\displaystyle\frac{N}{a}\displaystyle\frac{dF}{da}\displaystyle\frac{d\varphi}{da}\theta\bar\theta.
\end{equation}

This can be seen by the following explicit calculation 
\begin{equation}
\{\Omega,\widetilde\varphi\}=-N\{\widetilde{H},\varphi(a)\}\theta+\theta\displaystyle\frac{N}{a}\pi\displaystyle\frac{dF}{da}\displaystyle\frac{d\varphi}{da}-\displaystyle\frac{1}{a}\displaystyle\frac{dF}{da}\displaystyle\frac{d\varphi}{da}\rho\theta\bar\theta+\displaystyle\frac{1}{a}\displaystyle\frac{dF}{da}\displaystyle\frac{d\varphi}{da}\rho\theta\bar\theta=-N\{\mathcal{H},\varphi(a)\}\theta.
\end{equation}

Henceforth, the requirement (\ref{repr}) does not fix uniquely an element within each cohomological class $H^1(\Omega)$, which are formed by 
\begin{equation}
\Phi_{cc}=N\{\phi(a)\}\theta,\label{cc}
\end{equation}

$\{\phi\}$ being the equivalence class of functions under the Poisson action of the superHamiltonian $\mathcal{H}$, {\it i.e.}
\begin{equation} 
[\phi]=\{\phi'(a)| \phi'(a)=\phi(a)+\{\mathcal{H},\varphi(a)\}, \forall\varphi\}.\label{eqc}
\end{equation}

Therefore, the BRST-cohomological classses (\ref{cc}) are determined by the equivalence class of functions $\phi(a)$ under the Poisson action of $\mathcal{H}$. Hence they are in 1-1 correspondence with the classical observables of the FRW model (\ref{obsfrw}).

The Hamiltonian flow of the functions (\ref{cc}) in extended phase space is generated by the extended Hamiltonian (\ref{ham}). The Poisson bracket with $\Phi_{cc}$ gives
\begin{equation}
\{H,\Phi_{cc}\}=N\{ \widetilde{H},N[\phi(a)]\}\theta-[\phi(a)]\rho,
\end{equation} 
 
and one can take it back to a element $\Phi_{rep}$ by summing $\Theta=\{\Omega,\varphi\}$ with 
\begin{equation}
\varphi=\int^N \phi(a)dN'=N[\phi(a)].
\end{equation}

This way one obtains
\begin{eqnarray}
\{H,\Phi_{cc}\}_{rep}=\{H,\Phi_{cc}\}+\{\Omega,N\phi(a)\}\nonumber\\
=\{H,\Phi_{cc}\}-N\{\widetilde{H},N[\phi(a)]\}+[\phi(a)]\rho=0.
\end{eqnarray}

Therefore, the BRST-cohomological classses (\ref{cc}) do not evolve under the action of the physical Hamiltonian in extended phase-space.  

\section{Quantization in extended phase-space}\label{s4bis}
The quantization of the FRW model in extended phase space can be done as in the BFV case by defining proper states and a scalar product. Let us consider as quantum states the BRST-closed functions (\ref{rinv}) (which we denote by $\Phi$) and let us fix the operator ordering with momenta to the right. The complex structure is the standard complex conjugation, while ghosts $\theta$ and $\rho$ are real. The momenta are defined as $-i$ times the (left) derivatives operators of the corresponding variables.

We look for the extension of the scalar product (\ref{sp}), {\it i.e.}
\begin{equation} 
\langle\Phi_1,\Phi_2\rangle=\int \mu(a,N)da dNd\theta d\rho (N\phi_1)^*\theta  e^{i[\hat{K},\hat{\Omega}]}N\phi_2 \theta,\label{spfrw}
\end{equation}

where $\mu(a,N)$ is an un-specified measure. We cannot reproduce exactly the procedure described in section (\ref{sec1}) because of the nontrivial mixing of gauge and physical degrees of freedom. Let us keep for moment $\hat{K}$ as a generic operator having ghost number $-1$, {\it i.e.} 
\begin{equation}
\hat{K}=\hat{k}_1\bar\rho+\hat{k}_2\bar\theta,
\end{equation}

$\hat{k}_1$, $\hat{k}_2$ being two arbitrary operators. The operator $[\hat{K},\hat{\Omega}]$ acts on states (\ref{rinv}) as follows
\begin{equation}
[\hat{K},\hat{\Omega}]\Phi=-\hat{\Omega} \hat{K} \Phi= i\hat{\Omega}(\hat{k}_1 N\phi)=-iN\hat{\widetilde{H}}(\hat{k}_1N\phi)\theta-\frac{\partial}{\partial N}(\hat{k}_1 N\phi)\rho. \end{equation}
It is worth noting how the last term in the expression above is able to provide a nonvanishing result for the scalar product (\ref{spfrw}). By squaring the operator $[\hat{K},\hat{\Omega}]$ one gets
\begin{equation}
[\hat{K},\hat{\Omega}]^2\Phi=\hat{\Omega} \hat{K} \hat{\Omega} \hat{K}\Phi=\hat{\Omega}(\hat{k}_1N\hat{\widetilde{H}}(\hat{k}_1 N\phi)-i\hat{k}_2\frac{\partial}{\partial N}(\hat{k}_1 N\phi)).
\end{equation}

Let us make now some assumptions about the operators $\hat{k}_1$ and $\hat{k}_2$: for simplicity we take $\hat{k}_1=-Ii$, while we fix $\hat{k}_2$ as follows
\begin{equation}
\hat{k}_2=-N[\hat{\widetilde{H}},N]|_{\pi=0}-\frac{i}{2}N[[\hat{\widetilde{H}},N],N]\pi,\label{k2} 
\end{equation} 
where the first term is obtained by evaluating $-N[\hat{\widetilde{H}},N]$ and by avoiding the piece containing the operator $\hat{\pi}$. This way, the following relations hold  
\begin{eqnarray}
[\hat{K},\hat{\Omega}]\Phi=\hat{\Omega}(N\phi)=-N\hat{\widetilde{H}}(N\phi)\theta+i\frac{\partial}{\partial N}(N\phi)\rho,\label{ko1}\\
([\hat{K},\hat{\Omega}])^2\Phi=-\hat{\Omega}(N^2\hat{\mathcal{H}}\phi)=N\hat{\widetilde{H}}(N^2\hat{\mathcal{H}}\phi)\theta-i\frac{\partial}{\partial N}(N^2\hat{\mathcal{H}}\phi),\label{ko2}
\end{eqnarray}

and generically one has (see the appendix)
\begin{equation}
[\hat{K},\hat{\Omega}]^n\Phi=(-)^{n-1}\hat{\Omega}(N^n\hat{\mathcal{H}}^{n-1}\phi)=(-)^{n}N\hat{\widetilde{H}}(N^n\hat{\mathcal{H}}^{n-1}\phi)\theta+(-)^{n-1}i\frac{\partial}{\partial N}\left(N^n\hat{\mathcal{H}}^{n-1}\phi\right)\rho.
\end{equation}

Therefore, the exponential within the scalar product (\ref{spfrw}) reads
\begin{equation}
e^{i[\hat{K},\hat{\Omega}]}\Phi=\sum_n\frac{i^n}{n!}(-)^{n-1}\left(-N\hat{\widetilde{H}}(N^n\hat{\mathcal{H}}^{n-1}\phi)\theta+inN^{n-1}\hat{\mathcal{H}}^{n-1}\phi\rho\right),
\end{equation}
and after performing the integration over $\theta$ and $\rho$ only the second term gives a nonvanishing contribution so getting
\begin{equation}
\langle\Phi_1|\Phi_2 \rangle=-\int dadN \mu(a,N)\phi_1^*N\sum_n\frac{(-i)^{n-1}}{(n-1)!}N^{n-1}\hat{\mathcal{H}}^{n-1}\phi_2=-\int dadN \mu(a,N)\phi_1^*Ne^{-iN\hat{\mathcal{H}}}\phi_2.
\end{equation}
By defining the measure $\mu(a,N)=-\mu(a)/N$ in order to avoid the factor $N$ coming from $\Phi_1$, the scalar product above can be written as in (\ref{spf}), {\it i.e.}  
\begin{equation}
\langle\Phi_1|\Phi_2 \rangle=\int dadN \mu(a)\phi_1^*e^{iN\hat{\mathcal{H}}}\phi=\int da \mu(a)\phi_1^*\phi_2\delta(\hat{\mathcal{H}}).  \label{fin}
\end{equation}

Therefore, it is obtained the same Hilbert space structure as in the case the Dirac prescription for the quantization of constrained systems is used. 
In fact, there is no restriction on the form of the measure term $\mu(a)$ and one can write it as a $\delta$-function over a gauge-fixing condition times the proper factors which ensure the invariance under the choice of the gauge-fixing function itself. Hence, the expression (\ref{fin}) is the starting point to define a proper scalar product for the FRW model just like the case in which the Dirac prescription for the quantization of constrained systems is used \cite{ht}. 


\section{Conclusions}\label{s5}

In this work we derived the BRST charge associated with FRW space-time by a Noether-like analysis in extended phase space. The total Lagrangian has been defined according to the BV method for differential gauge fixing conditions. These conditions allowed us to reintroduce missing velocities and to have a well-defined Hamiltonian formulation in extended phase space \cite{Tatyana3}. Nihilpotent BRST transformations were defined thanks to the invariance under time parametrizations of the original formulation. The final expression for the action has been analyzed, finding that its variation under BRST transformations provided some time-boundary contributions. We accounted for these contributions and we could achieve the expression of the BRST charge for the considered systems. 

Then, we characterized the BRST cohomological classes in the case of functions with ghost number one. We chose proper elements within each BRST orbit, such that the closure condition fixied the dependence from $N$. This is the counterpart of the imposition of primary constraints in the Dirac approach. Then, the construction of equivalence class of closed forms modulo exact ones ensured the invariance under the action of the secondary constraint. We also investigate the physical evolution of cohomological classes under the action of the total Hamiltonian. We found that such an action vanished. This achievement confirms that the frozen formalisms extends to observables in extended phase space. Hence, the BRST formulation identifies the right degrees of freedom even in the case of a gravitational system, in which there is a nontrivial interplay between gauge (the lapse function) and physical (the scale factor) degrees of freedom (which is reflected into the expression of the BRST charge (\ref{brstfrw})).   
   
Finally, we considered the quantization of the FRW model in extended phase space. We demonstrated how a suitable scalar product could be defined according with the procedure described in \cite{ht} (the only difference being the form of the function $K$) such that the vanishing of the superHamiltonian operator was implemented. This achievement outlines how the equivalence between a quantum formulation in extended phase space and other approaches to the quantization of constrained systems, as refined algebraic quantization and the Dirac prescription, can be realized. Moreover, our findings can be thought as the canonical counterpart of the outcomes of \cite{Hall}, where a path integral formulation for a FRW model with a differential gauge fixing condition is discussed and the restriction to propagators implementing in a proper way the condition $\mathcal{H}=0$ is obtained. 

The extension of this analysis to more complex space-times will be the subject of forthcoming investigations, aimed to test to what extend the BRST framework can be used to implement diffeomorphisms invariance on a quantum level. The extension of the procedure to infer the conserved charge will require more algebraic manipulations with respect to the FRW case, since the Lagrangian in extended phase-space is more complex and the variation of the gravitational Lagrangian (which here vanishes on-shell) can give some additional contributions. However, such a Lagrangian will be obtained by discarding some boundary contributions from the Einstein-Hilbert one, whose associated action is invariant under space-time transformations. Therefore, even though the variation of the gravitational Lagrangian will contribute to the total variation of the action, the additional terms will still take the form of some boundary contributions and the definition of the conserved charge can be given as in Eq.(\ref{omega}). A different apporach is discussed in \cite{Tat}, where it is chosen to work with a Lagrangian containing second-order derivatives such that the whole action is invariant and the conserved charge is simply the Noether one.

\section*{Acknowledgment}   
The authors wish to thank the anonymous referees, whose remarks allowed them to enhanced the quality of the paper.\\ 
The work of F.C. was supported by funds provided by ``Angelo Della Riccia'' foundation and by the National Science Center under the agreement DEC-2011/02/A/ST2/00294.
This work has been realized in the framework of the CGW collaboration (www.cgwcollaboration.it).

\appendix 

\section*{Appendix}\label{app}

Let us demonstrate the relation
\begin{equation}
[\hat{K},\hat{\Omega}]^n\Phi=(-)^{n-1}\hat{\Omega}(N^n\hat{\mathcal{H}}^{n-1}\phi),\label{kon}
\end{equation}
which we have already verified for $n=1,2$ (\ref{ko1}), (\ref{ko2}). Let us assume that Eq.(\ref{kon}) holds and let us evaluate
\begin{eqnarray}
[\hat{K},\hat{\Omega}]^{n+1}\Phi=(-)^{n}\hat{\Omega} \hat{K}\hat{\Omega}\left(N^n\hat{\mathcal{H}}^{n-1}\phi\right)=\nonumber\\
=(-)^{n}\hat{\Omega} \hat{K}\left(-N\hat{\widetilde{H}}N^n\hat{\mathcal{H}}^{n-1}\phi\theta+i\frac{\partial}{\partial N}\left(N^n\hat{\mathcal{H}}^{n-1}\phi\right)\rho\right)=\nonumber\\
=(-)^{n}\hat{\Omega}\left(N\hat{\widetilde{H}}N^n\hat{\mathcal{H}}^{n-1}\phi+n\hat{k}_2(N^{n-1}\hat{\mathcal{H}}^{n-1}\phi)\right)=\nonumber\\
=(-)^{n}\hat{\Omega}\left(N^{n+1}\hat{\mathcal{H}}^{n}\phi+N[\hat{\widetilde{H}},N^n]\hat{\mathcal{H}}^{n-1}\phi+n\hat{k}_2(N^{n-1}\hat{\mathcal{H}}^{n-1}\phi)\right).\label{conto}
\end{eqnarray}

By using the expression (\ref{k2}) for $\hat{k}_2$ one finds
\begin{eqnarray}
\hat{k}_2(N^{n-1}\hat{\mathcal{H}}^{n-1}\phi)=(-N[\hat{\widetilde{H}},N]|_{\pi=0}-\frac{i}{2}N[[\hat{\widetilde{H}},N],N]\pi)(N^{n-1}\hat{\mathcal{H}}^{n-1}\phi)=\nonumber\\=-N^n[\hat{\widetilde{H}},N]\hat{\mathcal{H}}^{n-1}\phi-\frac{(n-1)}{2}N[[\hat{\widetilde{H}},N],N]N^{n-2}\hat{\mathcal{H}}^{n-1}\phi=\nonumber\\=-N^n[\hat{\widetilde{H}},N]\hat{\mathcal{H}}^{n-1}\phi-\frac{(n-1)}{2}N^{n-1}[[\hat{\widetilde{H}},N],N]\hat{\mathcal{H}}^{n-1}\phi,\label{conto1}
\end{eqnarray}
where in the last line we used the fact that the expression (\ref{tildeh}) for $\hat{\widetilde{H}}$ contains powers of $\pi$ up to the second order, thus the operator $[[\hat{\widetilde{H}},N],N]$ commutes with $N$. The second term in the last line of Eq.(\ref{conto}) contains the following object
\begin{eqnarray}
N[\hat{\widetilde{H}},N^n]=N\sum_{l=0}^{n-1}N^l[\hat{\widetilde{H}},N]N^{n-l-1}=\nonumber\\
=nN^{n}[\hat{\widetilde{H}},N]+N\sum_{l=0}^{n-1}N^l[[\hat{\widetilde{H}},N],N^{n-l-1}]=\nonumber\\
=nN^{n}[\hat{\widetilde{H}},N]+N\sum_{l=0}^{n-1}\sum_{m=0}^{n-l-2}N^lN^m[[\hat{\widetilde{H}},N],N]N^{n-l-2-m}=\nonumber\\=nN^{n}[\hat{\widetilde{H}},N]+N\sum_{l=0}^{n-1}\sum_{m=0}^{n-l-2}N^{n-2}[[\hat{\widetilde{H}},N],N]=\nonumber\\=nN^{n}[\hat{\widetilde{H}},N]+\frac{n(n-1)}{2}N^{n-1}[[\hat{\widetilde{H}},N],N],\label{conto2}
\end{eqnarray} 
where in the third line we still used the fact that $[[\hat{\widetilde{H}},N],N]$ commutes with $N$. By collecting togheter the results (\ref{conto1}) and (\ref{conto2}) one sees that
\begin{equation}
N[\hat{\widetilde{H}},N^n]\hat{\mathcal{H}}^{n-1}\phi+n\hat{k}_2(N^{n-1}\hat{\mathcal{H}}^{n-1}\phi)=0,
\end{equation}
and
\begin{equation}
[\hat{K},\hat{\Omega}]^{n+1}\Phi=(-)^{n}\hat{\Omega}(N^{n+1}\hat{\mathcal{H}}^{n}\phi),
\end{equation}
which probes eq.(\ref{kon}).

\end{document}